\documentclass[aps,floatfix,superscriptaddress]{revtex4}
\usepackage{amssymb,amsmath,amsfonts,latexsym,graphicx,epsfig,fancybox}
\usepackage{color}
\usepackage{titlesec}
\titleformat{\section}{\large\bfseries}{\thesection}{1em}{}

\newcommand{\bea}{\begin{eqnarray}}
\newcommand{\ena}{\end{eqnarray}}
\newcommand{\be}{\begin{equation}}
\newcommand{\en}{\end{equation}}
\newcommand{\nn}{\nonumber\\}

\newcommand{\Tr}{\mbox{\rm{tr}}}
\newcommand{\ed}{\end{document}}

\begin{document}

\hfill MITP/18-127 (Mainz)

\title{Ab initio three-loop calculation of the W-exchange contribution \\
to nonleptonic decays of double charm baryons} 

\author{Thomas Gutsche}
\affiliation{Institut f\"ur Theoretische Physik, Universit\"at T\"ubingen,
Kepler Center for Astro and Particle Physics, 
Auf der Morgenstelle 14, D-72076 T\"ubingen, Germany}
\author{Mikhail~A.~Ivanov}
\affiliation{Bogoliubov Laboratory of Theoretical Physics,
Joint Institute for Nuclear Research, 141980 Dubna, Russia}
\author{J\"urgen~G.~K\"orner}
\affiliation{PRISMA Cluster of Excellence, Institut f\"{u}r Physik,
Johannes Gutenberg-Universit\"{a}t, D-55099 Mainz, Germany}
\author{Valery E. Lyubovitskij}
\affiliation{Institut f\"ur Theoretische Physik, Universit\"at T\"ubingen,
Kepler Center for Astro and Particle Physics,
Auf der Morgenstelle 14, D-72076 T\"ubingen, Germany}
\affiliation{Departamento de F\'\i sica y Centro Cient\'\i fico
Tecnol\'ogico de Valpara\'\i so-CCTVal, Universidad T\'ecnica
Federico Santa Mar\'\i a, Casilla 110-V, Valpara\'\i so, Chile}
\affiliation{Department of Physics, Tomsk State University,
634050 Tomsk, Russia}
\affiliation{Laboratory of Particle Physics, Tomsk Polytechnic University,
634050 Tomsk, Russia}
\author{Zhomart Tyulemissov}
\affiliation{Bogoliubov Laboratory of Theoretical Physics,
Joint Institute for Nuclear Research, 141980 Dubna, Russia}

\begin{abstract}

We have made an ab initio three-loop quark model calculation of the 
$W$-exchange contribution to the nonleptonic two-body decays of the 
doubly charmed baryons $\Xi_{cc}^{++}$ and $\Omega_{cc}^{+}$. 
The $W$-exchange contributions appear in addition to the factorizable 
tree graph contributions and are not suppressed in general. 
We make use of the covariant confined quark model previously developed  
by us to calculate the tree graph as well as the $W$-exchange contribution.  
We calculate helicity amplitudes and quantitatively compare the tree graph 
and $W$-exchange contributions. Finally, we compare the calculated decay 
widths with those from other theoretical approaches when they are available. 

\end{abstract}

\maketitle

\section{Introduction}

The discovery of the double charm baryon state $\Xi_{cc}^{++}$ by
the LHCb Collaboration~\cite{Aaij:2017ueg} in the multibody decay mode
$(\Lambda_c K^- \pi^+ \pi^+)$ has provided a strong incentive for
further theoretical analysis of the weak decays of double charm baryons.
The lifetime of the $\Xi_{cc}^{++}$ has been measured to be
$(0.256^{+0.024}_{-0.022} ({\rm stat}) \pm 0.014 ({\rm syst}))$ 
ps~\cite{Aaij:2018wzf}.
The existence of the $\Xi_{cc}^{++}$ was confirmed in Ref.~\cite{Aaij:2018gfl},
again by the LHCb Collaboration, who reported on the first observation 
of a two-body nonleptonic decay of the doubly charmed baryon 
$\Xi_{cc}^{++}\rightarrow \Xi_{c}^{+}+\pi^{+}$. In the same report
the mass of the $\Xi_{cc}^{++}$ measured in~\cite{Aaij:2017ueg} 
was confirmed.  

The nonleptonic two-body decays of baryons have five different color-flavor
quark topologies. The set of contributing topological quark diagrams divides
into two groups: (i) the reducible tree-diagrams, 
and (ii) the irreducible  $W$--exchange diagrams.
The tree-diagrams are factorized into
the lepton decay of the emitted meson and the baryon-baryon transition
matrix elements of the weak currents. The $W$--exchange diagrams are more
difficult to evaluate from first principles. First attempts to estimate the
$W$--exchange contributions have been made in~\cite{Sharma:2017txj,Dhir:2018twm}
using a pole model approach and in~\cite{Jiang:2018oak} using final state
interactions based on triangle diagrams describing one-particle exchanges.
The authors
of~\cite{Sharma:2017txj,Dhir:2018twm} and~\cite{Jiang:2018oak} emphasize
that their results provide only first estimates of the $W$--exchange
contributions, in particular since their calculations involve generous 
approximations the errors of which are hard to quantify.

From the work of~\cite{Sharma:2017txj,Dhir:2018twm} one knows that the 
$W$--exchange contributions to nonleptonic double charm baryon decays are 
sizeable and cannot be neglected. The $W$--exchange contributions can 
interfere destructively or constructively with the tree diagram 
contributions.   
It is therefore of utmost importance to get the $W$--exchange contributions 
right. In this paper we set out to calculate the $W$--exchange
contributions to the Cabibbo favored nonleptonic two-body decays of
double charm baryons. We use the framework of our previously developed 
covariant constituent quark model to calculate the contributing 
three-loop quark Feynman diagrams. In a precursor of our present model 
some of us have calculated nonleptonic charm and bottom baryons including 
$W$--exchange contributions~\cite{Ivanov:1997ra}. 
We used a structureless static approximation for the light quark $(u,d,s)$ 
propagators and the leading-order contribution for the heavy quark $(c,b)$ 
propagators in the $1/m_{c/b}$ expansion. In the present calculation we use 
full quark propagators for the light and heavy quarks. We also now include 
quark confinement in an effective way. 

\section{Decay topologies of Cabibbo favored doubly charmed baryon 
nonleptonic decays}

\begin{figure}[ht]
\begin{center}
\epsfig{figure=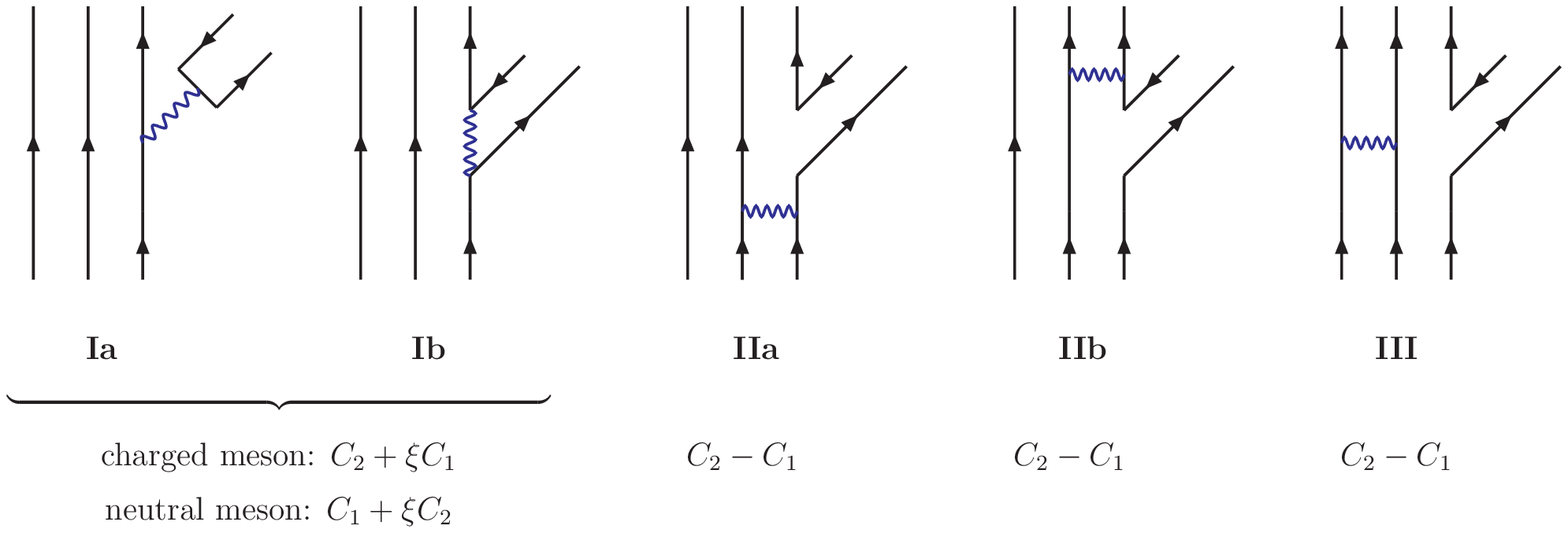,scale=.65}
\vspace*{-.25cm}
\caption{Flavor-color topologies of nonleptonic weak decays.}
\label{fig:NLWD}
\end{center}
\end{figure}

We begin by a discussion of the different color-flavor topologies that
contribute to the nonleptonic two-body transitions of the
double heavy $\Xi_{cc}$ and $\Omega_{cc}$ states. 
The relevant topologies are displayed in Fig.~\ref{fig:NLWD}. We refer to
the topologies of Ia and Ib as tree diagrams. They are also sometimes called
external (Ia) and internal $W$--emission (Ib) diagrams.
The topologies IIa, IIb, and III are referred to as $W$--exchange diagrams.
The labeling of the topologies follows the labeling introduced
in~\cite{Korner:1992wi,Korner:1994nh}. In \cite{Leibovich:2003tw} 
the $W$--exchange diagrams are denoted as the exchange (IIa), 
color-commensurate (IIb) and bow tie (III) diagram. The contribution of the
various topological diagrams to a particular decay is determined by the 
quark flavor composition of the particles involved in the decay.
For example, the decay 
$\Xi_{cc}^{++} \to \Sigma_c^{++} +\bar K^{\ast\,0}$ proceeds solely via the
tree diagram Ib. In~\cite{Gutsche:2017hux,Yu:2017zst} this decay has been
interpreted as making up a large part of the discovery final state channel
$(\Lambda_c^+ K^- \pi^+\pi^+)$ via the decay chain
$\Xi_{cc}^{++} \to \Sigma_c^{++}(\to \Lambda_c^+ +\pi^+) +\bar K^{\ast\,0}
(\to  K^- +\pi^+)$.

As shown in Fig.~\ref{fig:NLWD}, the color-flavor factor of the tree
diagrams Ia and Ib depend on whether the emitted meson is charged or neutral.
For charged emission the color-flavor factor is given by the 
combination of the Wilson coefficients $(C_2 + \xi C_1)$, 
where $\xi=1/N_c$ and $N_c$ is the number of colors, 
while for neutral emission the color-flavor factor
reads $(C_1 + \xi C_2)$. We take  $C_1=-0.51$ and $C_2=1.20$ at
$\mu=m_c=1.3$~GeV from Ref.~\cite{Buchalla:1995vs}. We use the large $N_c$
limit for  the color-flavor factors. For the $W$--exchange diagrams
the color-flavor factor is given by $(C_2-C_1)$. 

The decay of present interest
$\Xi_{cc}^{++} \to  \Xi_{c}^{+}  + \pi^+$~\cite{Aaij:2018gfl}
is fed by the tree diagram Ia and the $W$--exchange diagram IIb. We treat
this decay as well as the seven remaining $1/2^+ \to 1/2^+ + P(V)$ decays
that belong to the same topology class,
namely
\bea
\Xi_{cc}^{++} &\to&  \Xi_{c}^{+}  +\pi^+(\rho^+)\nn
\Xi_{cc}^{++} &\to&  \Xi_{c}^{\prime +}  + \pi^+(\rho^+)\nn
\Omega_{cc}^{+} &\to& \Xi_{c}^{+} + \bar K^0( K^{\ast\,0})\nn
\Omega_{cc}^{+} &\to& \Xi_{c}^{\prime +} + \bar K^0( K^{\ast\,0})
\ena
The $W$--exchange contributions to these decays fall into two classes.
The first class of these decays involves a $\Xi_{c}^{\prime +}$-baryon 
containing a symmetric $\{us\}$ diquark described by the interpolating current
$\varepsilon_{abc}\,(u^b C\gamma_\mu s^c)$, where $C = \gamma^0 \gamma^2$
is the charge conjugation matrix defined in terms of the Dirac matrices. 
The $W$--exchange contribution is strongly suppressed due to 
the K\"orner, Pati, Woo (KPW) theorem~\cite{Korner:1970xq,Pati:1970fg}. 
This theorem states that the contraction
of the flavor antisymmetric current-current operator with a flavor symmetric
final state configuration is zero in the $SU(3)$ limit. The antisymmetric
$[us]$ diquark emerging from the weak vertex is in the $3^*$ representation 
and cannot evolve into the $6$ representation of the symmetric final state 
$\{us\}$ diquark. In the following we will calculate $SU(3)$ breaking effects 
for the $W$--exchange contributions to this class of decays.
The second class involves a $\Xi_{c}^{+}$-baryon containing
a antisymmetric $[us]$ diquark described by the interpolating current
$\varepsilon_{abc}\,(u^b C\gamma_5 s^c)$. 
In this case the $W$--exchange contribution is not a priori suppressed. 
In Table~\ref{tab:charm} we display the quantum numbers, mass values, and
interpolating currents of double and single charmed baryons needed in 
this paper.  
  
\begin{table}[htb]
\caption{Quantum numbers and interpolating currents of double and
      single charmed baryons.} 
\centering
\def\arraystretch{1}
\begin{tabular}{l|c|c|c}
\hline
Baryon\qquad &\quad $J^P$ \quad & \quad Interpolating current \qquad &
\quad Mass (MeV)
\\
\hline 
$\Xi_{cc}^{++}$\quad \quad \quad & $\frac12^+$ &
$\varepsilon_{abc}\,\gamma^\mu\gamma_5 \, u^a (c^b C\gamma_\mu c^c)$  & 3620.6 
\\
$\Omega_{cc}^{+}$\quad \quad \quad & $\frac12^+$ &
$\varepsilon_{abc}\,\gamma^\mu\gamma_5 \, s^a (c^b C\gamma_\mu c^c)$ & 3710.0
\\
\hline
$\Xi_{c}^{'+}$ & $\frac12^+$ &
$\varepsilon_{abc}\,\gamma^\mu\gamma_5 \, c^a (u^bC\gamma_\mu s^c$) & 2577.4
\\
$\Xi_c^+$ & $\frac12^+$ & $\varepsilon_{abc}\, c^a (u^b C\gamma_5 s^c) $ & 2467.9 
\\
\hline
\end{tabular} 
\label{tab:charm}
  \end{table}

\section{Matrix elements and decay widths}

The effective Hamiltonian describing the $\bar s c\to \bar u d$ transition
is given by 
\bea
\mathcal {H}_{\rm eff} & = &
- g_{\rm eff} \, 
\left( C_1\,\mathcal{Q}_1 + C_2\,\mathcal{Q}_2\right)\, + \, {\rm H.c.},
\nn
\mathcal{Q}_1 &=&  (\bar s_a O_L c_b)(\bar u_b O_L d_a)
= (\bar s_a O_L d_a)(\bar u_b O_L c_b),
\nn
\mathcal{Q}_2 &=&  (\bar s_a O_L c_a)(\bar u_b O_L d_b)
= (\bar s_a O_L d_b)(\bar u_b O_L c_a),
\label{eq:eff-Ham}
\ena
where we use the notation $g_{\rm eff} = \frac{G_F}{\sqrt{2}} 
V_{cs} V^\dagger_{ud}$ and 
$O^\mu_{L/R}=\gamma^\mu(1 \mp \gamma_5)$
for the weak matrices with left/right chirality.

The nonlocal version of the interpolating currents shown in
Table~\ref{tab:charm} reads
\bea
J_{B_{cc}}(x)  &=& \int\!\! dx_1 \!\! \int\!\! dx_2 \!\! \int\!\! dx_3 \,
F_{B_{cc}}(x;x_1,x_2,x_3) \,
\varepsilon_{a_1a_2a_3}\,\gamma^\mu\gamma_5\, q_{a_1}(x_1)\,
\left(c_{a_2}(x_2) \,C\gamma_\mu \, c_{a_3}(x_3)\right)\,,
\nn
J_{B_c}(x)  &=& \int\!\! dx_1 \!\! \int\!\! dx_2 \!\! \int\!\! dx_3 \,
F_{B_c}(x;x_1,x_2,x_3) \,
\varepsilon_{a_1a_2a_3}\,\Gamma_1\, c_{a_1}(x_1)\,
\left(u_{a_2}(x_2) \,C\Gamma_2 \, s_{a_3}(x_3)\right)\,,
\nn
F_B &=& \delta^{(4)}\Big(x-\sum\limits_{i=1}^3 w_i x_i\Big)
\Phi_B\Big(\sum\limits_{i<j}(x_i-x_j)^2\Big) \,,
\label{eq:cur}
\ena
where $q=s$ or $u$, $w_i=m_i/(\sum\limits_{j=1}^3 m_j)$ and
$m_i$ is the quark mass at the space-time point $x_i$, and $\Gamma_1,\Gamma_2$
are the Dirac strings of the initial and final baryon states as specified
in Table~\ref{tab:charm}. 
Here $F_B$ and $\Phi_B$ are the Bethe-Salpeter kernel specifying the 
coupling of baryon with constituent quarks and correlation function, 
describing the distribution of quarks in baryon, respectively. 

\begin{figure}[ht]
\begin{center}
\epsfig{figure=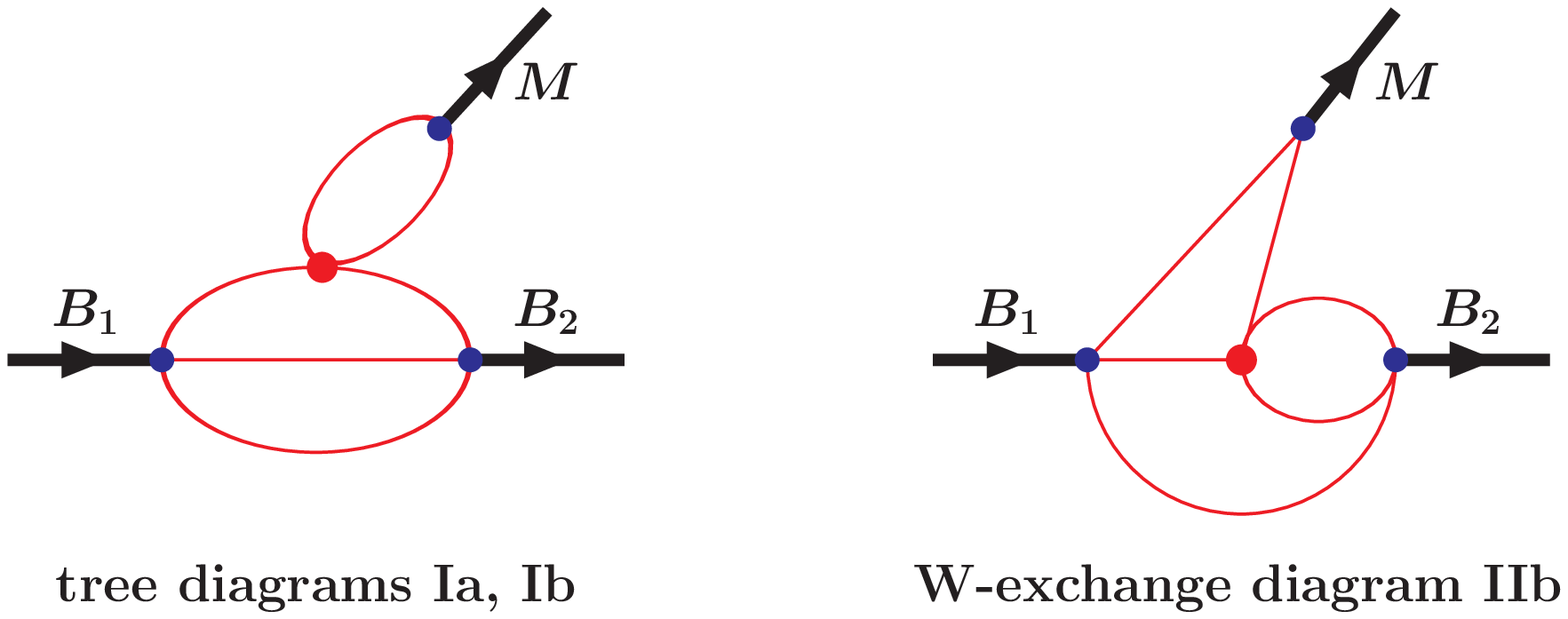,scale=.5}
\caption{Pictorial representations of Eqs.~(\ref{eq:tree})
  and (\ref{eq:W}).}
\label{fig:diag}
\end{center}
\end{figure}

The tree diagram and the IIb $W$--exchange contributions to the matrix element 
of the nonleptonic decays of the $\Xi_{cc}^{++}$ and $\Omega_{cc}^+$ read 
\bea
<B_2\,M|{\cal H}_{\rm eff}|B_1>
&=&
g_{\rm eff}\,\bar u(p_2)\Big( 12\,C_T\,M_T + 12\, (C_1-C_2)\,M_W \Big)u(p_1). 
\label{eq:matr_elem}
\ena
The tree diagram color factor for the neutral $\Omega_{cc}^+$ decays is
given by $C_T=-(C_1+\xi C_2)$ and by $C_T= + (C_2+\xi C_1)$ for the
charged $\Xi_{cc}^+$ decays.
The factor of $\xi=1/N_c$ is set to zero in our numerical calculations.
The overall factor of 12 in Eq.~(\ref{eq:matr_elem}) has its origin in a
combinatorial factor
of 2 and a factor of 6 from the contraction of two Levi-Civita color tensors.
The Feynman diagrams 
describing these processes are depicted in Fig.~\ref{fig:diag}.

\vspace*{-.2cm}
The contribution from the tree diagram factorizes into two pieces
according to
\bea
M_T &=& M_T^{(1)}\cdot M_T^{(2)},
\nn
M_T^{(1)} &=&
N_c\,g_{M}\int\!\!\frac{d^4k}{(2\pi)^4i}\widetilde\Phi_{M}(-k^2)\,
\Tr\left[O^\delta_L S_d(k-w_d q) \Gamma_{M} S_{s(u)}(k+w_{s(u)} q)\right]
\nn
M_T^{(2)}&=&
g_{B_1}g_{B_2}\int\!\!\frac{d^4k_1}{(2\pi)^4i}\int\!\!\frac{d^4k_2}{(2\pi)^4i}
\widetilde\Phi_{B_1}\Big(-\vec\Omega_1^{\,2}\Big)
\widetilde\Phi_{B_2}\Big(-\vec\Omega_2^{\,2}\Big)
\nn
&\times&
\Gamma_1 S_c(k_2)\gamma^\mu S_c(k_1-p_1) O_{R\,\delta} S_{u(s)}(k_1-p_2)
\widetilde\Gamma_2 S_{s(u)}(k_1-k_2)\gamma_\mu\gamma_5 \,.
\label{eq:tree}
\ena
Here $\Gamma_1\otimes\widetilde\Gamma_2=+I\otimes\gamma_5$ for
the $\Xi_c^{+}$-baryon and $-\gamma_\nu\gamma_5\otimes\gamma^\nu$
for the $\Xi_c^{\prime\,+}$-baryon.

The coupling constants $g_{M}$, $g_{B_1}$ and  $g_{B_2}$ are determined
as described in our previous papers (for details see, 
e.g.~\cite{Gutsche:2017hux,Gutsche:2018utw}). 
The Dirac matrix $\Gamma_{M}$ in $M_T^{(1)}$
reads $\gamma_5$ and $\epsilon_V\cdot\gamma$ for the pseudoscalar meson $P$
and for the vector meson $V$.
The connection of $M_T^{(1)}$ with the leptonic decay constants $f_M=f_P,f_V$
is given by $M_T^{(1)} = -f_P\,q^\delta$ and $+f_V m_V \,\epsilon_V^\delta$ . 
The minus sign  in front of $f_P$ appears because
the momentum $q$ flows in the opposite direction from 
the decay of $P$-meson. 
The Fourier-transforms of the vertex functions described by
the nonlocal interpolating currents are denoted by $\widetilde\Phi_H$.
We use for them the Gaussian functional form: 
$\widetilde\Phi_H(-k^2) = \exp(k^2/\Lambda^2_H)$, 
where $\Lambda_H$ is the hadron size parameter. 
Here and in the following the arguments of the baryonic vertex functions are 
expressed through the Jacobi momenta $(q_1,q_2)$ and $(r_1,r_2)$ by 
$\vec\Omega_1^{\,2} = \tfrac12 (q_1+q_2)^2 + \tfrac16 (q_1-q_2)^2, 
\ \vec\Omega_2^{\,2} =  \tfrac12 (r_1+r_2)^2 +\tfrac16 (r_1-r_2)^2\,.$  
The momenta $q_i$ and $r_i$ are defined from momenta conservation 
in each vertex of the diagrams 
(see details in Ref.~\cite{Gutsche:2017hux,Gutsche:2018utw}). 

The calculation of the three-loop $W$--exchange contribution 
is much more involved because the matrix element does not factorize.
By using the Fierz transformation
$O_{L/R}^{\alpha_1\alpha_2}O_{R/L}^{\alpha_3\alpha_4}=
2\,(1\pm\gamma_5)^{\alpha_1\alpha_4} (1\mp\gamma_5)^{\alpha_3\alpha_2}$ 
one has 
\bea 
\vspace*{-.1cm}
M_W &=&
g_{B_1}g_{B_2}g_{M}
\int\!\!\frac{d^4k_1}{(2\pi)^4i}
\int\!\!\frac{d^4k_2}{(2\pi)^4i}
\int\!\!\frac{d^4k_3}{(2\pi)^4i}
\widetilde\Phi_{B_1}\Big(-\vec\Omega_1^{\,2}\Big)
\widetilde\Phi_{B_2}\Big(-\vec\Omega_2^{\,2}\Big)
\widetilde\Phi_{M}(-P^2)\,
\nn
&\times&
2\,\Gamma_1\, S_c(k_1)\gamma^\mu S_c(k_2)(1-\gamma_5)S_d(k_2-k_1+p_2)
\Gamma_{M}S_{s(u)}(k_2-k_1+p_1)\gamma_\mu\gamma_5
\nn
&\times&
\Tr\Big[S_{u(s)}(k_3)\widetilde\Gamma_2 S_{s(u)}(k_3-k_1+p_2)(1+\gamma_5)\Big]\,,
\label{eq:W}
\ena
where $\Gamma_1\otimes\widetilde\Gamma_2=I\otimes\gamma_5$ for  
$B_2=\Xi^{+}_c$ and $-\gamma_\nu\gamma_5\otimes\gamma^\nu$  
for $B_2=\Xi^{\prime\,+}_c$. Here $P = k_2-k_1+w_d\,p_1 + w_u\,p_2$ 
is the Jacobi momentum in the meson vertex function.  

We are now in the position to verify the KPW theorem in our three-loop
calculation. To do this, we change the order
of Dirac matrices in the trace by using the properties of the charge
conjugation matrix. Keeping in mind that $\gamma_5$ does not contribute
to the trace, we have 
\vspace*{-.1cm}
\bea
\Tr\Big[S_u(k_3)\gamma_\nu S_s(k_3-k_1+p_2)]
=-\,\Tr\Big[S_s(-k_3+k_1-p_2)\gamma_\nu S_u(-k_3)\Big]\,.
\label{eq:trace}
\ena 
We insert Eq.~(\ref{eq:trace}) into  Eq.~(\ref{eq:W}) and shift
the integration variable $k_3\to -k_3+k_1-p_2$. One can check that
$\vec\Omega_2^{\,2}$ goes into itself under this transformation
accompanied by an interchange
of the $u-$ and $s-$ quark masses.
Thus, if $m_u=m_s$ then $M_W$ is identical zero which directly confirms
the KPW--theorem. We have checked numerically that the three-loop integral
vanishes in this limit.

Details of the calculation of the loop integrals and the subsequent reduction
of the integration over Fock-Schwinger variables to an integration
over a hypercube may be found in our previous papers (see e.g. the most recent
papers~\cite{Gutsche:2017hux,Gutsche:2018utw}). 
Compared to the two-loop calculation
of~\cite{Gutsche:2017hux,Gutsche:2018utw}) we are now dealing with 
a three-loop calculation involving  six quark propagators instead of 
the four propagators in the two-loop case.
The calculation is quite time-consuming both analytically and numerically.

Next one expands the transition amplitudes in terms of invariant amplitudes.
One has
\vspace*{-.1cm}
\bea
&&
<B_2\,P|{\cal H}_{\rm eff}|B_1>
=
g_{\rm eff}\,
\bar u(p_2)\left( A+\gamma_5\,B \right)u(p_1)\,,
\label{eq:ampl-P}
\\
&&
<B_2\,V|{\cal H}_{\rm eff}|B_1>
=
g_{\rm eff}\,
\bar u(p_2)\,\epsilon^\ast_{V \delta}
\left( \gamma^\delta\,V_{\gamma}+p_1^\delta\,V_{p}
+\gamma_5 \gamma^\delta\,V_{5\gamma}+\gamma_5 p_1^\delta\,V_{5p} \right)
u(p_1).
\label{eq:ampl-V}
  \ena
The invariant amplitudes are converted to a set of helicity amplitudes
$H_{\lambda_1\,\lambda_M}$ as described in~\cite{Korner:1992wi}. One has  
 \bea
 H^V_{\tfrac12\,t} &=& \sqrt{Q_+}\,A\,,\qquad
 H^A_{\tfrac12\,t} = \sqrt{Q_-}\,B\,,
 \nn[-2mm]
 H^V_{\tfrac12\,0} &=&
 + \sqrt{Q_-/q^2}\,\Big( m_+\, V_\gamma + \tfrac12 Q_+\,V_p\Big)\,,
 \,\,\,\,\qquad
 H^V_{\tfrac12\,1}  = -\sqrt{2Q_-}\,V_\gamma\,, 
 \nn[-2mm]
 H^A_{\tfrac12\,0} &=&
+ \sqrt{Q_+/q^2}\,\Big( m_-\, V_{5\gamma} + \tfrac12 Q_-\,V_{5p}  \Big)\,,
 \qquad
 H^A_{\tfrac12\,1}  = - \sqrt{2Q_+}\,V_{5\gamma}\,, 
 \ena 
where $m_\pm=m_1\pm m_2$, $Q_{\pm}=m_\pm^2-q^2$ and 
$\mathbf{|p_2|}=\lambda^{1/2}(m_1^2,m_2^2,q^2)/(2m_1)$.   
The helicities of the three particles are related by  
$\lambda_1 = \lambda_2 - \lambda_M$. We use the notation 
$\lambda_P=\lambda_t=0$ for the scalar $(J=0)$ contribution 
in order to set the helicity label apart from $\lambda_V=0$ 
used for the longitudinal component of the $J=1$ vector meson. 
The remaining helicity amplitudes can be obtained from the parity relations
$H^V_{-\lambda_2,-\lambda_M} = + H^V_{\lambda_2,\lambda_M}$ and 
$H^A_{-\lambda_2,-\lambda_M} = - H^A_{\lambda_2,\lambda_M}$\,. 
The helicity amplitudes have the dimension $[m]^3$. 
The numerical results on the helicity
amplitudes given in Tables~\ref{tab:OmS}-\ref{tab:XiA} are in units of GeV$^3$.

The two-body decay widths read
\vspace*{-.2cm}
  \bea
  \Gamma(B_1\to B_2+P) &=& 
  \frac{g_{\rm eff}^2}{16\pi}\frac{\mathbf{|p_2|}}{m_1^2}\,
  {\mathcal H}_S\,, \quad 
  {\mathcal H}_S =  
  \Big|H_{ \tfrac12\,t}\Big|^2 \,+\, 
  \Big|H_{-\tfrac12\,t}\Big|^2 \,, 
  \label{eq:width_P}
  \\[-2mm]
  \Gamma(B_1\to B_2+V) &=& 
  \frac{g_{\rm eff}^2}{16\pi}\frac{\mathbf{|p_2|}}{m_1^2}\,
  {\mathcal H}_V\,, \quad 
  {\mathcal H}_V = 
  \Big|H_{ \tfrac12\, 0}\Big|^2 \,+\,  
  \Big|H_{-\tfrac12\, 0}\Big|^2 \,+\, 
  \Big|H_{ \tfrac12\, 1}\Big|^2 \,+\, 
  \Big|H_{-\tfrac12\,-1}\Big|^2 \,, 
  \label{eq:width_V}
 \ena
where we denote the sum of the squared moduli of the  helicity amplitudes
$H=H^V-H^A$ by ${\cal H}_S$ and ${\cal H}_V$~\cite{Gutsche:2018utw}. 

\section{Numerical results}

All model parameters have been fixed in our previous studies except for
the size parameter $\Lambda_{cc}$ of the double charmed baryons. As a first
approximation we equate the size parameter of double charm baryons with
that of single charm baryons, i.e. we take
$\Lambda_{cc}=\Lambda_{c}=0.8675$~GeV where we
adopt the value of $\Lambda_{c}$ from~\cite{Gutsche:2015rrt}. 
Numerical results for the helicity amplitudes and decay widths
are displayed in the Tables~\ref{tab:OmS}-\ref{tab:XiA}.
In this paper we concentrate on our predictions for rate values.
On top of the rate predictions, Tables~\ref{tab:OmS}-\ref{tab:XiA} contain a
wealth of spin polarization information. For example, for the decay
$\Xi_{cc}^{++}\rightarrow \Xi_{c}^{+}+\pi^{+}$ one finds an asymmetry
parameter of $\alpha = - 2 H^V_{1/2\,0}H^A_{1/2\,0}/(|H^V_{1/2\,0}|^2
+|H^A_{1/2\,0}|^2)=-0.57 $ while~\cite{Sharma:2017txj} predict a value
in the range $\alpha= [-0.86, -1.00]$ depending on their model assumptions.
Note that the $W$--exchange contribution in~\cite{Sharma:2017txj} is purely
$p$--wave, i.e. proportional to $H^A_{1/2\,0}$,
due to the nonrelativistic approximations that they
employ. This is in stark contrast to our relativistic result where the
$s$--wave amplitude dominates in this process,
i.e. $H^V_{1/2\,0}/H^A_{1/2\,0}=3.3$. Both model calculations agree on a very
substantial destructive interference of the tree and $W$--exchange
contributions.

Our results highlight the importance of the KPW theorem for the nonleptonic
decays when the final state involves a $\Xi^{\prime+}$ baryon containing a
symmetric
$\{su\}$ diquark. Tables~\ref{tab:OmS}-\ref{tab:XiA} show that the relevant
$W$--exchange contributions are strongly suppressed. Nonzero values result
from $SU(3)$ breaking effects which are accounted for in our approach.
Take for example the decay $\Xi_{cc}^{++} \to \Xi_c^{'+} +\pi^+$. When
compared to the tree contribution the $SU(3)$ breaking effects amount
to $\sim(2-4)\,\%$. While the consequences of the KPW theorem for the
$W$--exchange contribution are incorporated in the pole model approach
of~\cite{Sharma:2017txj} they are not included in the final-state interaction
approach of~\cite{Jiang:2018oak}.

In Table~\ref{tab:comparison} we compare our rate results with the
results of some other 
approaches~\cite{Dhir:2018twm,Sharma:2017txj,Jiang:2018oak,%
Wang:2017mqp,Yu:2017zst,Kiselev:2001fw}. Note that the rates calculated
in~\cite{Wang:2017mqp} include tree graph contributions only. There is a wide
spread in the rate values predicted by the various model calculations. All
calculations approximately agree on the rate of the decay
$\Xi_{cc}^{++} \to \Xi_c^{'+} +\rho^+$ which is predicted 
to have a large branching
ratio of $\sim 16\, \%$. In our calculation this mode is predicted to have
by far the largest branching ratio of the decays analyzed in this paper.
As concerns the decay $\Xi_{cc}^{++}\rightarrow \Xi_{c}^{+}+\pi^{+}$ discovered
by the LHCb Collaboration~\cite{Aaij:2018gfl} we find a branching ratio of
${\cal B}(\Xi_{cc}^{++}\rightarrow \Xi_{c}^{+}\pi^{+})=0.70 \,\%$
using the central value of the life time measurement in~\cite{Aaij:2018wzf}.  
The small value of the branching ratio results from a substantial 
cancellation of the tree and $W$--exchange contributions. 
The branching ratio is somewhat smaller than the branching ratio 
${\cal B }(\Xi_{cc}^{++} \to \Sigma_c^{++} + \bar K^0)=1.28\,\%$ calculated 
in~\cite{Gutsche:2017hux}. 
We predict a branching ratio considerably smaller than the range of
branching fractions $(6.66 - 15.79)\,\%$ calculated in~\cite{Sharma:2017txj}.

An important issue is the accuracy of our results. The only free parameter
in our approach is the size parameter $\Lambda_{cc}$ of the double heavy
baryons for which 
we have chosen $\Lambda_{cc}=0.8675$~GeV in Tables~\ref{tab:OmS}-\ref{tab:XiA}.
In order to estimate the uncertainty caused by the choice of the
size parameter we allow the size parameter to vary from 0.6 to 1.135~GeV.
We evaluate the mean 
$\bar\Gamma = \sum\Gamma_i/N$ and the mean square deviation
$\sigma^2=\sum (\Gamma_i-\bar\Gamma)^2/N$. The results for $N=5$
are shown in Table~\ref{tab:errors}. The rate errors amount to $6 - 15 \%$.
Since the dependence of the rates on $\Lambda_{cc}$ is nonlinear the central
values of the rates in Table~\ref{tab:errors} do not agree
with the rate values in~Tables~\ref{tab:OmS}-\ref{tab:XiA}.

\begin{table}[!htbp]
\vspace*{-.7cm}
\begin{tabular}{cc}
    \begin{minipage}{.45\linewidth}
   \centering
   \caption{
      $\Omega^+_{cc}\to\Xi^{\prime\,+}_{c} + \bar K^0(\bar K^{\ast\,0})$}
    \label{tab:OmS}
    \vskip 1 mm
     \def\arraystretch{.9}
\begin{tabular}{|cccc|}
\hline
  Helicity        & Tree diagram &  $W$ diagram & total  \\
\hline 
$ H^V_{\tfrac12\,t} $ &  $0.20$     & $-0.01$   & $0.19$  \\ 
$ H^A_{\tfrac12\,t} $ &  $0.25$     & $-0.01$   & $0.24$
\\[1.1ex]
\hline
\multicolumn{4}{|c|}
            {$\Gamma(\Omega^+_{cc}\to\Xi^{\prime\,+}_{c}+ \bar K^0)
              = 0.15\cdot 10^{-13}\,\text{GeV}\ $} \\
\hline 
$ H^V_{\tfrac12\,0} $ &  $-0.25$    & $0.04 \times 10^{-1}$    &  $-0.25$ \\ 
$ H^A_{\tfrac12\,0} $ &  $-0.50$    & $ 0.01$    &  $-0.49$ \\ 
$ H^V_{\tfrac12\,1} $ &  $ 0.27$    & $-0.01$    &  $ 0.26$ \\ 
$ H^A_{\tfrac12\,1} $ &  $ 0.56$    & $0.04 \times 10^{-2}$    &  $ 0.56$
\\[1.1ex] 
\hline
\multicolumn{4}{|c|}
            {$\Gamma(\Omega^+_{cc}\to\Xi^{\prime\,+}_{c}+ \bar K^{\ast\,0})
              = 0.74 \cdot 10^{-13}\,\text{GeV}\ $} \\
\hline
\end{tabular} 
\end{minipage} &

    \begin{minipage}{.45\linewidth}

   \centering
   \caption{
      $\Omega^+_{cc}\to\Xi^{+}_{c} + \bar K^0(\bar K^{\ast\,0})$ }
    \label{tab:OmA}
    \vskip 1 mm
     \def\arraystretch{.9}
\begin{tabular}{|cccc|}
\hline
  Helicity        & Tree diagram &  $W$ diagram & total  \\
\hline 
$ H^V_{\tfrac12\,t} $ &  $-0.35$     & $1.06$   & $0.71$  \\ 
$ H^A_{\tfrac12\,t} $ &  $-0.10$     & $0.31$    & $0.21$
\\[1.1ex]
\hline
\multicolumn{4}{|c|}
            {$\Gamma(\Omega^+_{cc}\to\Xi^{+}_{c}+ \bar K^0)
              = 0.95 \cdot 10^{-13}\,\text{GeV}\ $} \\
\hline 
$ H^V_{\tfrac12\,0} $ &  $ 0.50$   & $-0.69$   &  $-0.19$ \\ 
$ H^A_{\tfrac12\,0} $ &  $ 0.18$   & $-0.45$   &  $-0.27$ \\ 
$ H^V_{\tfrac12\,1} $ &  $-0.11$   & $-0.24$   &  $-0.35$ \\
$ H^A_{\tfrac12\,1} $ &  $-0.18$  & $ 0.66$    &  $ 0.48$
\\[1.1ex]
\hline
\multicolumn{4}{|c|}
            {$\Gamma(\Omega^+_{cc}\to\Xi^{+}_{c}+ \bar K^{\ast\,0})
              = 0.62\cdot 10^{-13}\,\text{GeV}\ $} \\
\hline
\end{tabular} 
\end{minipage} 
\end{tabular}

\begin{tabular}{cc}
    \begin{minipage}{.45\linewidth}
   \centering
   \caption{
      $\Xi^{++}_{cc}\to\Xi^{\prime\,+}_{c} + \pi^+(\rho^+)$ }
    \label{tab:XiS}
    \vskip 1 mm
     \def\arraystretch{.9}
\begin{tabular}{|cccc|}
\hline
  Helicity      & Tree diagram &  $W$ diagram & total  \\
\hline 
$ H^V_{\tfrac12\,t} $ &  $-0.38$     & $-0.01$ & $-0.39$  \\ 
$ H^A_{\tfrac12\,t} $ &  $-0.55$     & $-0.02$ & $-0.57$
\\[1.1ex]
\hline
\multicolumn{4}{|c|}
            {$\Gamma(\Xi^{++}_{cc}\to\Xi^{\prime\,+}_{c} + \pi^+)
              = 0.82\cdot 10^{-13}\,\text{GeV}\ $} \\
\hline 
$ H^V_{\tfrac12\,0} $ &  $ 0.60$   & $0.04 \times 10^{-1}$   &  $0.61$ \\ 
$ H^A_{\tfrac12\,0} $ &  $ 1.20$   & $0.01$   &  $1.21 $ \\ 
$ H^V_{\tfrac12\,1} $ &  $-0.49$   & $-0.01$  &  $-0.50$ \\ 
$ H^A_{\tfrac12\,1} $ &  $-1.27$   & $0.01 \times 10^{-1}$ &  $-1.27$ 
\\[1.1ex] 
\hline
\multicolumn{4}{|c|}
            {$\Gamma(\Xi^{++}_{cc}\to\Xi^{\prime\,+}_{c} + \rho^+)
              = 4.27\cdot 10^{-13}\,\text{GeV}\ $} \\
\hline
\end{tabular} 
\end{minipage} &

    \begin{minipage}{.45\linewidth}

   \centering
   \caption{
      $\Xi^{++}_{cc}\to\Xi^{+}_{c} + \pi^+(\rho^+)$}
    \label{tab:XiA}
     \vskip 1 mm
    \def\arraystretch{.9}
\begin{tabular}{|cccc|}
\hline
  Helicity         & Tree diagram &  $W$ diagram & total  \\
\hline 
$ H^V_{\tfrac12\,t} $ &  $-0.70$     & $0.99$ & $0.29$  \\ 
$ H^A_{\tfrac12\,t} $ &  $-0.21$     & $0.30$ & $0.09$
\\[1.1ex]
\hline
\multicolumn{4}{|c|}
            {$\Gamma(\Xi^{++}_{cc}\to\Xi^{+}_{c} + \pi^+)
              = 0.18\cdot 10^{-13}\,\text{GeV}\ $} \\
\hline
$ H^V_{\tfrac12\,0} $ &  $ 1.17$    & $-0.70$   &  $ 0.47$  \\
$ H^A_{\tfrac12\,0} $ &  $ 0.45$    & $-0.44$   &  $ 0.003$ \\
$ H^V_{\tfrac12\,1} $ &  $-0.20$    & $-0.23$   &  $-0.43$  \\
$ H^A_{\tfrac12\,1} $ &  $-0.41$    & $ 0.62$   &  $ 0.21$
\\[1.1ex]
\hline
\multicolumn{4}{|c|}
            {$\Gamma(\Xi^{++}_{cc}\to\Xi^{+}_{c} + \rho^+)
              = 0.63\cdot 10^{-13}\,\text{GeV}\ $} \\
\hline
\end{tabular} 
\end{minipage}
\end{tabular}

   \vskip 1mm 
   \centering
   \caption{Comparison with other approaches. Abbreviation: M=NRQM, T=HQET}
    \label{tab:comparison}
     \def\arraystretch{.9}
\begin{tabular}{|l|c|c|c|c|c|c|}
\hline
\qquad Mode &  \multicolumn{6}{|c|}{Width (in $10^{-13}$~GeV)} \\ 
\cline{2-7}
& \qquad  our\qquad\qquad & Dhir~\cite{Dhir:2018twm,Sharma:2017txj} 
& Jiang~\cite{Jiang:2018oak} & Wang~\cite{Wang:2017mqp} 
& Yu~\cite{Yu:2017zst}    &  Kiselev~\cite{Kiselev:2001fw}
\\
\hline 
$\Omega^+_{cc}\to\Xi^{\prime\,+}_{c} + \bar K^0$  & 0.15   & 0.31 (M)  & & & &
\\
                                                  &        & 0.59 (T)  & & & &
\\
\hline
$ \Omega^+_{cc}\to\Xi^{+}_{c}+ \bar K^0 $         & 0.95  & 0.68 (M) &  & & &
\\
                                                  &        & 1.08 (T) &  & & &
\\
\hline
$\Omega^+_{cc}\to\Xi^{\prime\,+}_{c} + \bar K^{\ast\,0}$ & 0.74  
& & $2.64^{+2.72}_{-1.79}$ &  & &
\\
\hline
$\Omega^+_{cc}\to\Xi^{+}_{c}+ \bar K^{\ast\,0}$& 0.62 & & $1.38^{+1.49}_{-0.95}$ 
&  & &
\\
\hline
$\Xi^{++}_{cc}\to\Xi^{\prime\,+}_{c} + \pi^+$  & 0.82 &1.40 (M) &   & 1.10 & &
\\
                                               & & 1.93 (T)&      &        & &
\\
\hline
$\Xi^{++}_{cc}\to\Xi^{+}_{c} + \pi^+$          & 0.18 &1.71 (M) &   & 1.57 
& 1.58 & 2.25
\\
                                               & & 2.39 (T)&   & & &
\\
\hline
$\Xi^{++}_{cc}\to\Xi^{\prime\,+}_{c} + \rho^+$ & 4.27 & & $4.25^{+0.32}_{-0.19}$  
& 4.12 & 3.82 &
\\
\hline
$\Xi^{++}_{cc}\to\Xi^{+}_{c} + \rho^+$         & 0.63 & & $4.11^{+1.37}_{-0.86}$  
& 3.03 & 2.76 & 6.70 
\\
\hline
\end{tabular} 

\vskip 1mm 
\centering
   \caption{Estimating uncertainties in the decay widths.}
    \label{tab:errors}
     \def\arraystretch{.825}
\begin{tabular}{|lc|}
  \hline
\qquad   Mode \qquad  & \qquad Width (in $10^{-13}$~GeV) \qquad \\
  \hline
  $\Omega^+_{cc}\to\Xi^{\prime\,+}_{c} + \bar K^0$ \qquad &
\qquad $0.14 \pm  0.01$ \qquad  
\\  
$\Omega^+_{cc}\to\Xi^{\prime\,+}_{c} + \bar K^{\ast\,0}$ \qquad &
\qquad $0.72 \pm  0.06$ \qquad 
\\
\hline
$\Omega^+_{cc}\to\Xi^{+}_{c} + \bar K^0$ \qquad   &
\qquad $0.87 \pm 0.13$ \qquad 
\\
$\Omega^+_{cc}\to\Xi^{+}_{c} + \bar K^{\ast\,0}$ \qquad &
\qquad $0.58 \pm 0.07$ \qquad 
\\
\hline
$\Xi^{++}_{cc}\to\Xi^{\prime\,+}_{c} + \pi^+$ \qquad  &
\qquad $0.77 \pm 0.05$ \qquad 
\\
$\Xi^{++}_{cc}\to\Xi^{\prime\,+}_{c} + \rho^+$ \qquad &
\qquad $4.08 \pm 0.29$ \qquad 
\\
\hline
$\Xi^{++}_{cc}\to\Xi^{+}_{c} + \pi^+$ \qquad &
\qquad $0.16 \pm 0.02$\qquad 
\\
$\Xi^{++}_{cc}\to\Xi^{+}_{c} + \rho^+$ \qquad &
\qquad $0.59 \pm 0.04$ \qquad
\\
\hline
\end{tabular} 
\end{table}

\section{Outlook}

We now have the tools at hand to calculate all Cabibbo favored and
Cabibbo suppressed nonleptonic two-body decays of the double charm
ground state baryons $\Xi_{cc}^{++}$, $\Xi_{cc}^{+}$, and $\Omega_{cc}^+$.
These would also include the $1/2^+ \to 3/2^+ + P(V)$ nonleptonic decays
not treated in this paper.
Of particular interest are the modes
$\Xi_{cc}^+ \to \Sigma^{(*)+} + D^{(*)0}\,({\rm III})$,
$\Xi_{cc}^+ \to \Xi^{(*)0} + D_s^{(*)+}\,({\rm III})$, and
$\Omega_{cc}^+ \to \Xi^{(*)0} + D^{(*)+}\,({\rm IIb})$
which are only fed by a single W-exchange contribution as indicated in
apprentices. Of these the three modes involving the final state
$3/2^+$ baryons $\Sigma^{*+}$
and $\Xi^{*0}$ would be forbidden due to the KPW theorem. It would be
interesting to check on this prediction of the quark model.

\vspace*{-.35cm}
\begin{acknowledgments}
\vspace*{-.3cm}

This work was funded by
the Carl Zeiss Foundation under Project ``Kepler Center f\"ur Astro-  
und Teilchenphysik: Hochsensitive Nachweistechnik zur Erforschung des 
unsichtbaren Universums (Gz: 0653-2.8/581/2)'',  
by CONICYT (Chile) PIA/Basal FB0821, by the Russian Federation  
program ``Nauka'' (Contract No. 0.1764.GZB.2017), 
by Tomsk State University competitiveness improvement  
program (Grant No.~8.1.07.2018), and by Tomsk Polytechnic University 
Competitiveness Enhancement Program (Grant No.~VIU-FTI-72/2017).  
M.A.I.\ acknowledges the support from the PRISMA Cluster of Excellence 
(Mainz Uni.).

\end{acknowledgments}

\end{document}